\documentclass[]{article}
\usepackage{graphics}
\usepackage[varg]{txfonts} 
\usepackage[latin1]{inputenc}
\begin{document}
\title{Simulations of drastically reduced SBS with laser pulses com\-posed of a Spike Train
of Uneven Duration and Delay (STUD pulses)}
\author{Stefan H\"uller\footnote{email: hueller@cpht.polytechnique.fr}\\
Centre de Physique Th\'eorique, CNRS, Ecole Polytechnique, Palaiseau, France\\
Bedros Afeyan \\
Polymath Research Inc., Pleasanton, CA, USA}
\maketitle
%
\abstract{
By comparing the impact of established laser smoothing techniques like Random Phase Plates (RPP) and
Smoothing by Spectral Dispersion (SSD) to the concept of ``Spike Trains of Uneven Duration and Delay''
(STUD pulses) on the amplification of parametric instabilities in laser-produced plasmas, we show with the help of numerical simulations, that STUD pulses can drastically reduce instability growth by orders of magnitude. 
The simulation results, obtained with the code {\sc Harmony} in a nonuniformly flowing mm-size plasma for the Stimulated Brillouin Scattering (SBS) instability, show that the efficiency of the STUD
pulse technique is due to the fact that successive re-amplification in space and time of parametrically
excited plasma waves inside laser hot spots is minimized.
An overall mean fluctuation level of ion acoustic waves at low amplitude is established because of the
frequent change of the speckle pattern in successive spikes. This level stays
orders of magnitude  below the levels of ion acoustic waves excited in hot spots of RPP and SSD laser beams.
} 
\vspace*{-.4cm}
\section{Introduction} \label{intro}
\vspace*{-.15cm}
The concept of "Spike Trains of Uneven Duration and Delay" (STUD pulses \cite{stud}) is a novel approach
that aims to overcome the problem strongly growing parametric instabilities in laser-produced plasmas in
the context of inertial confinement fusion (ICF) and high energy density physics (HEDP) more generally. The method was conceived in 2009 by B. Afeyan \cite{stud} and is explained in detail in a companion article by Afeyan \cite{ifsa-bedros} in this volume.  The physical implementation of STUD pulses in large laser systems offers the prospect of controlling the unbridled 
growth of electron plasma or ion acoustic waves driven by Raman or Brillouin scattering. 

So-called optical smoothing techniques, like Random Phase Plates (RPP), Smoothing by Spectral
Dispersion (SSD), and Induced Spatial Incoherence (ISI) have been developed in the last three decades to
diminish the risk of hydrodynamic instabilities in fusion capsules, as well as with the hopes that residually, they may limit the growth of
laser-plasma instabilities. The impact of these smoothing techniques on parametric instabilities, like stimulated Brillouin scattering, is the subject of the work presented here. Unfortunately, neither RPP nor SSD can efficiently inhibit growth of parametric instabilities inside potentially cooperative regimes between thousands of  laser-intensity "hot spots" that form the typical speckle patterns of smoothed laser beams.

"Spike Trains of Uneven Duration and Delay"  require the use of phase plates such as RPP (or "CPP"), where the laser beam spatial profiles are a speckle pattern.
However, two essential features distinguish STUD pulses from previous techniques\cite{ifsa-bedros}:

(i) instead of following a continuous profile, the laser pulse is subdivided in spikes in which
the intensity is partially "on" and partially "off". The average intensity over the spike duration,
however, equals the value equivalent to a continuous pulse, while the peak intensity of the
spike is increased by a factor that depends on the duty cycle;

(ii) the speckle patterns are scrambled in between spikes either each time or after a number of spikes.
If the scrambling is in between each spike, they are called STUD$\times$1 pulses.
In the case of STUD$\times$Inf (STUD$\times\infty$) pulses, a single RPP or speckle
pattern will be deployed during the entire pulse with no speckle platter changes.
Note that a STUD$\times$Inf pulse with 100\% duty cycle is equivalent to an RPP (or CPP) beam.

The principal idea behind STUD pulses is to control  parametrically
excited waves from getting re-amplified continuously, there where they have been excited previously.
The frequent change of speckle pattern in between spikes has the effect that re-amplification
of a plasma mode, generated in a hot spot, is very  unlikely due to the exponential statistics of the intensities of hot spots.
Our results validate the picture that the interaction volume tends towards being successively filled
with relatively low-level ion acoustic waves (IAW). In contrast to this, speckle patterns in RPP beams,
constantly amplify IAWs in situ, in particular in intense speckles. In case of
SSD we can speak of repeated re-amplification which is very deleterious.

Here we show the results of  STUD pulses with a 50\%
duty cycle (equal "on" and "off" time intervals, on average, and twice higher peak intensity with respect
to a continuous pulse).
We also introduce a random jitter to the width of the on pules which is 10\% in magnitude.
This is in order not to excite ion acoustic waves by the harmonics of the periodic picket fence skeleton of the pulse.
This gives rise to so-called STUD5010$\times$1 pulses. For details and other constructions, see \cite{ifsa-bedros}.
We will show the advantages of STUD5010$\times$1 over the performance of  STUD5010$\times$Inf pulses and the advantages of both of these over SSD and RPP beams.
\vspace*{-.5cm}
\section{Simulations with {\sc Harmony}}
\vspace*{-.2cm}
We have explored the performance of STUD pulses in the case of SBS.
On the acoustic spatial and temporal scales, it is possible to consider 
laser-plasma interactions in 2D that include thousands of laser speckles. This guarantees sufficiently
good speckle statistics and can be run for upwards of a 100 ps at moderate computational expense.
With the help of numerical simulations using the code {\sc Harmony} \cite{harmony},
we have compared the evolution of this instability to cases where the driving laser 
is RPP, SSD, and 50\% duty cycle STUD pulses with changing speckle patters with each spike,
STUD5010$\times$1, and with nonchanging speckle patterns, STUD5010$\times$Inf.
{\sc Harmony} solves two paraxial wave equations for the incident light field with the complex-valued 
amplitude $E_0$ (with principal frequency and wavenumber $\omega_0$, $k_0$) and for a backscattered light
field amplitude $E_1$ (with $\omega_1$, $k_1$) propagating in opposite direction to each other
($k_1 \simeq -k_0$).
In the version of {\sc Harmony} used here, the ion acoustic wave (IAW) relevant in the 3-wave coupling of
SBS is resolved by a set of 3 (scalar) equations for continuity and for the momenta in $z$ and $x_{\perp}$ of this short-wavelength component ($k_z \simeq 2 k_0$), 
(while in the version described in Ref. \cite{harmony}, a single envelope approach is used
for the SBS-driven IAW)
\begin{eqnarray}
&
\left( \partial_t + \textbf{u}_0\cdot \nabla \right) n_1 + n_0 \nabla\cdot \textbf{u}_1 + n_1 \nabla\cdot \textbf{u}_0 = 0 \quad \mbox{(1)}, \hspace{.2cm} \mbox{and} \hspace{.2cm} 
&\left( \partial_t + \textbf{u}_0\cdot \nabla  + 2 \nu_{\rm IAW} \right) \textbf{u}_1 + c_s^2 \nabla n_1 = \textbf{f}_{\rm SBS} \quad \mbox{(2)} \label{mom} \ ,   \nonumber
\end{eqnarray}
where $n_0$ and $\textbf{u}_0$ stand for the average density and for advective flow, respectively,
$\nu_{\rm IAW}$ for IAW damping, $c_s$ for the sound speed, and $\textbf{f}_{\rm SBS}$ for the
acceleration due to the SBS coupling with both light field components,
$\textbf{f}_{\rm SBS} \propto n_0 \nabla  E_0 E_1^* \exp(i 2 k_0 z) $.
Equations (1-2) for $n_1$ and $\textbf{u}_1 \equiv (u_{1,z},u_{1,x})$
are solved with the help of a conservative Eulerian scheme, with flux correction, for the
two vectors $U_{\pm} \equiv c_s n_1 \pm n_0 u_{z,1}$ for the
(complexed-valued) density and velocity perturbations,
$n_1$, $u_{1,z}$, and $u_{1,x}$.
The vectors $U_{\pm}$ follow the counter-propagating characteristics
of sound, as well as for the 3rd $x$-component of Eq. (2) for $c_s u_{1,x}$. 
The IAW density perturbation $n_1 = (U_+ +U_- )/2 c_s$ is then used for the 3-wave coupling terms for
the paraxial equations for $E_0$ and $E_1$ (see \cite{harmony}). The usual 1st order equation for $n_1$
for `weak coupling', when temporal growth is slower than the ion acoustic oscillations,
corresponds to the case when the counter-propagating wave $U_-$ is not excited.
Solving Eqs. (1) and (2) allow to access the regime of `strong coupling' of SBS
\cite{strong-coupling} that may easily appear to be relevant in high-intensity speckles of the smoothed
beam.
The dominant term for advection in the laser propagation direction in Eqs. (1-2)
corresponds to a frequency shift $\textbf{u}_0\cdot \nabla \simeq i 2 k_0 u_{0,z}$. 
A flow profile is used in our simulations with
$u_{z,0}(z) = u(L_z) [ 1 - (z-L_z)/u(L_z) L_u ]$.
In order to seed SBS growth of possibly unstable modes with frequency
$\sim 2 k_0 [c_s - u_{z,0}(z)]$, everywhere in the simulation box, a small-amplitude seed in
the field $E_1$ is introduced a the right-hand-side (RHS) boundary $z=L_z$,
having a broadband spectrum with $\delta\omega \sim 25 k_0 c_s$ around the laser frequcency $\omega_0$.
This is done via a Langevin equation $(\partial_t + \delta\omega ) E_1(L_z,t) = \delta\omega \ S(t)$ with
random noise generation each $\Delta t$, respecting $\delta\omega \Delta t \ll 1$.
The module for the 3-wave coupling, describing propagation of laser- and backscattered
light, and for the short-wavelength IAWs driven by backscatter SBS, solved via Eqs. (1)-(2).
is coupled to a nonlinear hydrodynamics module that describes both the plasma expansion
as well as the (long-wavelength) acoustic waves associated with forward-SBS and self-focusing,
Harmony takes into account the momentum transfer between the short- and
long-wavelength components, see herefore Ref. \cite{pesme_ppcf}.
The algorithm of {\sc Harmony} has the options to take into account, or to exclude,
(i) self-focusing and momentum transfer,
(ii) pump depletion,
and (iii) higher harmonics of the IAW.
\begin{figure}
\resizebox{0.6\columnwidth}{!}{%
\includegraphics{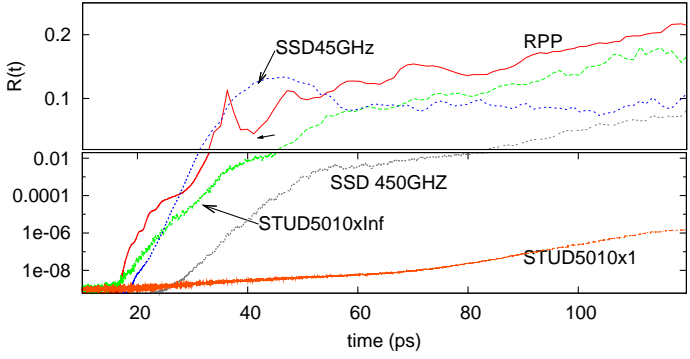}}
\caption{Reflectivity of SBS backscatter as a function of time from simulations with {\sc Harmony} for
different the different smoothing methods discussed here: RPP, SSD at 45GHz and 450GHz bandwidth, STUD5010$\times$, and STUD5010 with a fixed speckle pattern STUD5010$\times$Inf.
Note: upper/lower subplot in linear/logarithmic scale.}
\label{fig1} 
\end{figure}
The simulations whose results are shown here have been carried out in two-dimensional (2D)
geometry in a box of 4000 laser wavelengths, $\lambda_0,$ along the laser propagation direction (spatial resolution $\Delta z =2\lambda_0$) 
and 8192$\lambda_0$, in the transverse direction ($\Delta x =\lambda_0$), with the wavelength
chosen to be $\lambda_0 = 0.53 \mu$m.
For a laser $f_{\#}$-number of  $f_{\#}=$8, the box length $L_z$ contains $~\sim$9 effective hot spots
with an equal number of hot spot lengths in between them.
The effective hot spot length is $L_{\rm HS} = \pi f_{\#}^2 \lambda_0 \simeq 220\lambda_0$.
The non-periodic boundary conditions in $x$ result in an effective box-width of $\sim$4000$\lambda_0$ (about half the entire box width),
resulting in a number $>$1500 potential speckles, sufficiently numerous to avoid too strong a fluctuation
from realization to realization \cite{speckle-stat}.
In all of the simulations, a laser pulse with a flat intensity profile in $x$,
lasting 3 times the transit time of the box, $3 L_z/c$, precedes the main laser pulse.
This pulse is used to initialize equivalent conditions for the noise in the plasma for each of the
different types of main laser pulses that follow, either RPP, SSD, or STUD pulses.
The main pulse is smoothly switched on over $L_z/c$.
The simulated plasma at $T_e =$2 keV, had 10\% critcal density, $n_0 =$0.1$n_c$, with a inhomomeneous
flow profile in the interval $u_{0,z}/c_s =$-3$\ldots$-7$c_s$ with an inhomogeneity length (at $z=L_z$)
of $L_v = |u(L_z)| L_u =$3000$\lambda_0$. The IAW damping coefficient was
$(\nu/\omega)_{\rm IAW} =$0.025.
The corresponding SBS (Rosenbluth) gain for the average intensity of a RPP and SSD runs at 
$I_L =$5.5$\times$ 10$^{14}$W/cm$^2$ (and twice this value for STUD pulse peaks)
in the inhomogeneous profile hence was $G_{\rm SBS}=$ 5.5 at $z=L_z$.
and the interaction length $L_{\rm int} = 2 (\nu/\omega)_{\rm IAW} [c_s + u_{0,z}(z)] L_v /u_{0,z}(z)$
thus $L_{\rm int} =$ 100$\ldots$ 128$\lambda_0$.
For the applied STUD pulses the spike duration chosen was $c t_{\rm spike}=$173 $\lambda_0$.
For the pulses following SSD a bandwidth of 45GHz was chosen. An alternative simulation with a
10 times larger bandwidth at 450GHz was also carried out to see the impact of faster
smoothing, even if it lies outside any realistic modulation pace of SSD systems.
\begin{figure}
\resizebox{0.95\columnwidth}{!}{%
\includegraphics{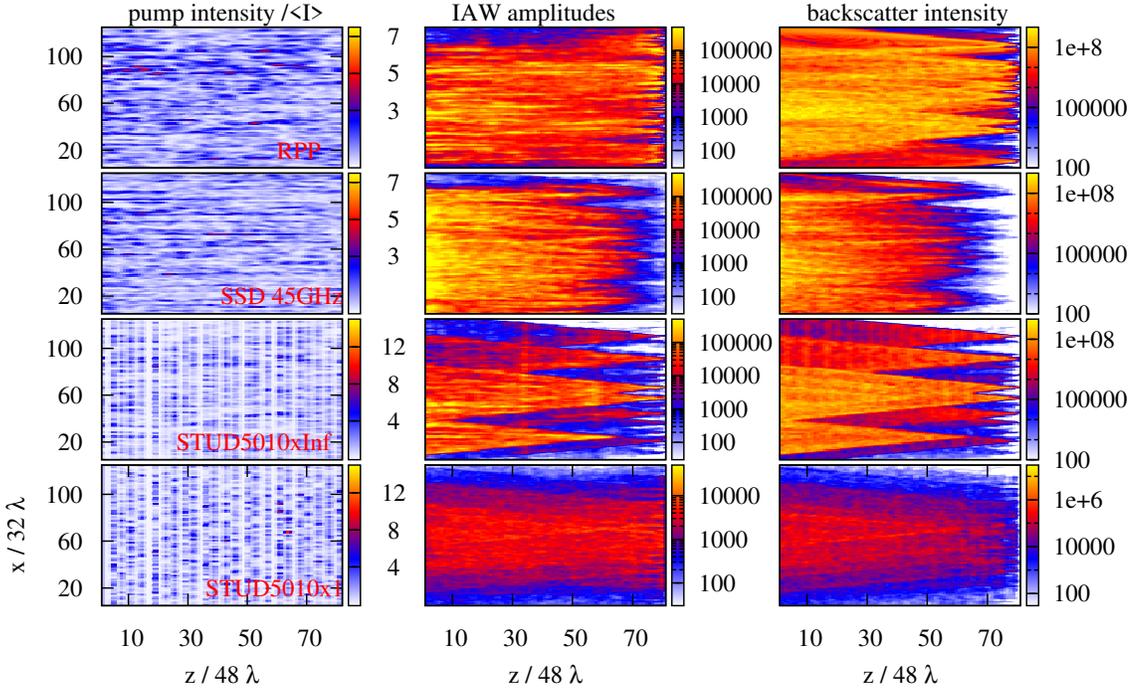} }
\caption{Snapshots in the $x-z$-plane, taken at $t=$100ps, of laser pump intensity (left column), of backscattered light intensity (center), and of IAW amplitudes (right),
for the cases (top to bottom) RPP, SSD at 45GHz, STUD5010$\times$Inf, STUD5010$\times$1 pulses.
The pump intensity is normalized to the beam average intensity $\langle I\rangle$,
both other quantities to the level after the pulse preceding the main pulse. 
Note the difference in the levels (see color bars) for the STUD5010$\times$1 case 
with respect to the other cases. RPP case (upper line):
two regions with particularly high IAW amplitudes (center) coinciding with intense speckle locations (left) are high-lighted in frames.
}
\label{fig2}
\end{figure}

\noindent
{\bf Simulation results:} The simulations with {\sc Harmony} have been carried out starting from a noise level of
$\sim 3 \times 10^{-9}$ of the incoming laser light pump intensity (of the late time RPP beam) 
in the backscattered light intensity.
As illustrated in Fig. \ref{fig1}, the cases for RPP and SSD at 45GHz rapidly attain pump
depletion in numerous speckles, attaining a SBS reflectivity around 10-20\%\cite{remark}.
SSD proves to show a reduction with respect to RPP on a longer time scale\cite{Berger},
but the efficiency of reduction is less than an order of magnitude for the realistic bandwidth
value of 45GHz. The reflectivity for the 10 times higher bandwidth value exhibits a slower growth,
but tends again to reflectivities above 1\%.
As can be seen from Fig. \ref{fig2}, the RPP case (and to a lower degree the SSD case)
exhibits high-amplitude IAW structures in space due to repeated amplification 
in close vicinity to high-intensity speckles (see zones put in frames). 
Apparently the SSD technique, that can be associated with
moving speckles within a limited volume,
cannot avoid that IAWs are re-amplified sufficiently often by intense speckles.
Compared to both the RPP and the SSD technique, the use of STUD pulses
results in a drastic reduction of SBS backscattering.
Figure \ref{fig2} furthermore illustrates that the mean level of IAW fluctuations for the
STUD pulse stays of at least 2 orders of magnitude below the
levels seen for all other cases. In particular, spatial structures that can be associated with the
speckle pattern are washed out.
It is also instructive to note that the STUD5010$\times$Inf pulse case, based on a fixed speckle pattern, but using
the the same spikes, attains similar levels as RPP and SSD \@ 45GHz. This underlines the fact that
a frequent scrambling of hot spots into uncorrelated speckle patterns is an important factor for the success of
the STUD pulse program.

Part of the simulations have be performed on the facilities of IDRIS-CNRS, Orsay, France.
Partial support from the programs DOE NNSA SSAA, Phase I SBIR from DOE OFES, and DOE NNSA-OFES Joint HEDP is acknowledged.


\begin{thebibliography}{}
\bibitem{remark} The values of both RPP and SSD cases based on a single speckle pattern
are subject to statistical fluctuations with respect to an ensemble average over realisations.
Presence of pump depletion in the simulations has reduced, however, the fluctuations.
\end{thebibliography}
\end{document}